\documentclass[aip,jcp,reprint]{revtex4-1}

\usepackage{amsmath}

\usepackage{graphicx}
\usepackage{url}
\usepackage{braket}
\usepackage{paralist}

\usepackage{xcolor}

\newcommand{\eref}[1]{Eq.~\ref{#1}}
\newcommand{\fref}[1]{Fig.~\ref{#1}}
\newcommand{\sref}[1]{Sec.~\ref{#1}}
\newcommand{\aref}[1]{Appendix~\ref{#1}}

\newcommand{\sgn}{\ensuremath{\text{sgn}}}

\begin{document}


\title{Minimising biases in Full Configuration Interaction Quantum Monte Carlo} 



\author{W. A. Vigor}
\affiliation{Department of Chemistry, Imperial College London, Exhibition Road, London, SW7 2AZ, United Kingdom}
\author{J. S. Spencer}
\affiliation{Department of Physics, Imperial College London, Exhibition Road, London, SW7 2AZ, United Kingdom}
\affiliation{Department of Materials, Imperial College London, Exhibition Road, London, SW7 2AZ, United Kingdom}
\author{M. J. Bearpark}
\affiliation{Department of Chemistry, Imperial College London, Exhibition Road, London, SW7 2AZ, United Kingdom}
\author{A. J. W. Thom}
\affiliation{Department of Chemistry, Imperial College London, Exhibition Road, London, SW7 2AZ, United Kingdom}
\affiliation{University Chemical Laboratory, Lensfield Road, Cambridge, CB2 1EW, United Kingdom}


\date{\today}

\begin{abstract}
We show that Full Configuration Interaction Quantum Monte Carlo (FCIQMC) is a Markov Chain in its present form. We construct the Markov matrix of FCIQMC for a two determinant system and hence compute the stationary distribution. These solutions are used to quantify the dependence of the population dynamics on the parameters defining the Markov chain.  Despite the simplicity of a system with only two determinants, it still reveals a population control bias inherent to the FCIQMC algorithm.  We investigate the effect of simulation parameters on the population control bias for the neon atom and suggest simulation setups to in general minimise the bias. We show a reweighting scheme to remove the bias caused by population control commonly used in Diffusion Monte Carlo [J. Chem. Phys. \textbf{99}, 2865 (1993)] is effective and recommend its use as a post processing step.
\end{abstract}


\maketitle 

\section{Introduction}


A cheap and accurate computational description of the ground state energy of a chemical system remains one of the principal challenges in electronic structure theory, yet achieving both of these goals systematically remains beyond the grasp of current approximations.  Hierarchies of methods of increasing sophistication have been developed in the quantum chemistry community which systematically capture increasing amounts of the electron-electron correlation energy at the expense of additional computational cost. 
These methods start from Hartree--Fock\cite{RevModPhys.23.69} which scales modestly with the fourth power of the number of electrons to Full Configuration Interaction (FCI) which captures the maximal amount of electron-electron correlation in a finite basis set but scales factorially with the number of electrons. Approximations (which are often very accurate) such as density fitting can potentially reduce the scaling of these methods.\cite{doi:10.1080/0026897042000274801} If FCI is used with a large enough basis set or an extrapolation to the complete basis set limit\cite{Halkier1998243,aipjournaljcp1011010.10631.468080}, energy differences can be obtained to chemical accuracy (1 kcal/mol) providing direct comparison with experiment. Unfortunately the factorial scaling with the number of electrons makes it unfeasible for studying anything but the smallest of chemical systems.

Full Configuration Interaction Quantum Monte Carlo (FCIQMC)\cite{booth:054106} marries FCI with a projector Monte Carlo paradigm but crucially requires no \emph{a priori} knowledge of the sign structure of the wavefunction.
FCIQMC has two principal advantages over conventional FCI:~the storage requirements are greatly reduced due to a sparse stochastic representation of the wavefunction\cite{booth:054106} and it can be efficiently parallelised.\cite{doi:10.1080/00268976.2013.877165}
The storage requirements for FCIQMC depend on the (system-dependent) severity of the Fermion sign problem\cite{spencer:054110} and are often orders of magnitude less than conventional FCI calculations.  FCIQMC with the controllable initiator approximation\cite{cleland:041103} has allowed molecular systems with Hilbert spaces of $10^{29}$ Slater determinants\cite{ct300486d} and the uniform electron gases with Hilbert spaces containing up to $10^{108}$ determinants\cite{PhysRevB.85.081103} to be studied.
The FCIQMC methodology was subsequently extended to coupled cluster\cite{PhysRevLett.105.263004}, and we believe stochastic approaches are becoming increasingly important to the quantum chemistry community due to the need for scalable algorithms which are well-suited to modern computer architectures.

Some questions still remain concerning the best way to use Monte Carlo to solve the FCI equations. The FCIQMC algorithm is not a black box, and a choice has to be made about calculation parameters which control the stochastic sampling and hence the systematic and stochastic errors inherent to the simulation for a given amount of computational resources.  In this article we investigate the behaviour of FCIQMC simulations to understand the relationship between parameter choices and errors by investigating the exact distribution obtained from a Markov Chain transition matrix.

\sref{fciqmc_recap} contains a brief recap of the FCIQMC method.  We show in \sref{stochastic_matrices} that FCIQMC is an example of Markov Chain Monte Carlo (MCMC).  We use these ideas to investigate population control bias in the two determinant H$_2$ system and more realistic calculations on the neon atom in \sref{finite_pop_bias}.  We draw conclusions and provide suggestions on simulation strategies in \sref{discussion}.  Atomic units are used throughout.  The many-electron Hamiltonian and all energies have been shifted to be relative to the absolute Hartree--Fock energies of the appropriate system.
Error bars signify one standard error, an estimate of the standard deviation, either side of
the mean value.

\section{FCIQMC}
\label{fciqmc_recap}

We briefly review the FCIQMC method, which is discussed in more detail in (e.g.) Refs.~\onlinecite{booth:054106,spencer:054110}.
The imaginary-time Schr\"odinger equation is
\begin{equation}
\frac{\partial \ket{\Psi}}{\partial \tau} = -(\hat{H} - S) \ket{\Psi},
\label{diffusion}
\end{equation}
where $S$ is an energy offset, which we shall discuss in the context of FCIQMC later, introduced to control normalisation.
The general solution to \eref{diffusion} is $\ket{\Psi(\tau)} = e^{-\tau(\hat{H}-S)} \ket{\Psi(\tau=0)}$, which in the long-time limit tends to the lowest eigenstate with which the initial wavefunction has a non-zero overlap.

We begin with the configuration interaction (CI) ansatz where the wavefunction is a linear combination of Slater determinants: $\ket{\psi} = \sum_i C_i \ket{D_i}$.
It is convenient (though not necessary\cite{PhysRevLett.109.230201}) to represent the coefficients by a discrete set of signed particles which we shall call \emph{psips}\cite{anderson:1499}.
Booth \emph{et al.}\cite{booth:054106} showed that a finite-difference approximation to \eref{diffusion} could be sampled by allowing a psip on one determinant to create a new psip on another determinant (`spawn') or on the same determinant (`death') with probability proportional to the connecting Hamiltonian matrix element.
Pairs of psips with opposite signs on the same determinant are removed (`annihilated') at the end of each timestep.
After a sufficient number of such steps, the psip vector becomes a stochastic representation of the eigenvector.
The finite difference approximation introduces no timestep errors if the timestep, $\delta\tau$, satisfies $\delta \tau < 2(E_\text{max}  - E_0)^{-1}$, where $E_\text{max}$ ($E_0$) is the highest (lowest) eigenvalue of the Hamiltonian\cite{spencer:054110}; a property FCIQMC shares with Green's function quantum Monte Carlo\cite{PhysRevB.41.4552}. 

Following an equilibration phase, the shift is periodically updated every $A$ steps to control the psip population using\cite{booth:054106}
\begin{equation}
	S(\tau + A \delta \tau) = S(\tau) -  \frac{\gamma}{A \delta \tau} \log{\frac{N(\tau +A\delta \tau)}{N(\tau)}},
\label{Shift}
\end{equation}
where $\gamma$ is a damping factor and $N(\tau)$ is the total number of psips at time $\tau$.
Repeated substitution of \eref{Shift} into itself yields:
\begin{equation}
    S(\tau + A \delta \tau) = S(0) - \xi \log{\frac{N(\tau +A\delta \tau)}{N_s}},
\label{shift-1}
\end{equation}
where $S(0)$ is the initial value of the shift (in this work the Hartree--Fock energy), $N_s$ is the population at the end of the equilibration phase and $\xi=\gamma/(A\delta\tau)$ is usually fixed during a simulation.
\eref{shift-1} implies that FCIQMC is an example of Markov Chain Monte Carlo (MCMC), the implications of which we shall discuss in the next section.

The correlation energy can also be found by:
\begin{equation}
E_{\text{Proj}} = \frac{\braket{D_0 | \hat{H} e^{-\hat{H}\tau} | D_0}}{\braket{D_0 | e^{-\hat{H}\tau} | D_0}} =  \frac{\sum_{i\ne 0} H_{0i} n_i}{n_0},
\end{equation}
where the trial state, $\ket{D_0}$, is typically the Hartree--Fock determinant.
The variance of the projected estimator is generally smaller than that of the shift and can be reduced further by a multi-determinant trial function.\cite{PhysRevLett.109.230201}

Both estimators are serially correlated as the state of the simulation at one timestep is heavily dependent on the state at the previous timestep.
We use an automated iterative blocking algorithm\cite{10.1063/1.457480,Wolff2004143,PhysRevE.83.066706,pyblock} to accurately estimate the stochastic error in all FCIQMC calculations presented in this paper.

\section{Stochastic Matrices and FCIQMC}
\label{stochastic_matrices}

\subsection{General Markov Chain Monte Carlo Theory}

A stochastic process is a discrete time Markov Chain if the probability of transitioning from one state to another (in one discrete timestep) depends only upon the current state.\cite{Sokal,kemeny1976finite}
The probability of a given set of psips, $\{n_i\}$, producing another set of psips, $\{n_i^{\prime}\}$, at the next timestep depends only on $\{n_i\}$ and the value of the shift, which depends upon the total number of psips at a given shift update.
We can therefore describe FCIQMC as a Markov chain taking one step every $A$ timesteps.  In order to simplify the mathematical details, we shall henceforth assume that the simulation takes one timestep between shift updates (i.e. $A=1$).

We shall denote the Markov states of the simulation using indices $\alpha,\beta \dots$ and Slater determinants using indices $i, j, \dots$.  The stochastic matrix, $\Gamma$, consists of elements $\Gamma_{\alpha,\beta}$ which give the probability that the system transitions from state $\alpha$ to state $\beta$ in one step in the Markov chain and is in general not symmetric.

We can infer some properties of the eigenvectors and eigenvalues of the stochastic matrix as $\Gamma$ is non-negative and $\sum_{\beta} \Gamma_{\alpha \beta} = 1$ 
as an FCIQMC calculation must transition from one state to another or remain in the same state.
From this there must exist one or more left eigenvectors, $\gamma_\alpha$, satisfying:
\begin{equation}
\sum_\alpha \gamma_\alpha\Gamma_{\alpha \beta}  = \gamma_{\beta},
\end{equation}
where $\gamma_{\alpha}$ gives the probability that the Markov chain will be in state $\alpha$ if the chain is in equilibrium. 
The Perron--Frobenius theorem proves that the $\Gamma$ must have one or more such Perron--Frobenius eigenvectors with a unit eigenvalue and all other eigenvalues must be smaller.
The Perron--Frobenius eigenvector is unique and the chain will converge towards this distribution if \begin{inparaenum}[\upshape (\itshape i\upshape )] \item all the states are aperiodic i.e. $\Gamma^N_{\alpha \beta} > 0$ for all values of large $N$; \item every state can be reached from every other state.\end{inparaenum}
For function $f(\alpha)$ defined for all possible Markov states, its expectation value is:
\begin{equation}
\mu_f = \langle f_t \rangle_t = \sum_\alpha f(\alpha)  \gamma_{\alpha}.
\end{equation}
The Perron--Frobenius eigenvector specifies the distribution of an ensemble of independent Markov chains taking a single step, and by computing it we may find expectation values of interest in this system.

\subsection{The FCIQMC Markov chain}

A state $\alpha$ in FCIQMC is represented by the signed number of psips on each determinant ($n_{a}, n_{b}, \dots$): 
\begin{equation}
    \alpha := (n_{a}^{(\alpha)}, n_{b}^{(\alpha)}, \dots ).
\end{equation}
The shift $S$ is not an independent variable as it is simply a function of the total number of psips. The FCIQMC chain is in an absorbing state (i.e. the probability of leaving the state is zero) when there are no psips on any of the determinants, as all events which change the psip population require an existing non-zero population.
As the shift is initially held at a constant value during the equilibration phase, the Markov chain changes after the shift is allowed to vary.

The estimators of interest in FCIQMC are the shift and the numerator and denominator of the projected energy:
\begin{gather}
    S(\alpha) = S(\tau=0) -\xi \log\frac{N(\alpha)}{N_s} \\ 
    E_\textrm{Numer}( \alpha) = \sum_{i\ne 0} H_{0 i} n_{i}^{(\alpha)} \\ N_\textrm{Denom}( \alpha) = n_{0}^{(\alpha)}.
\end{gather}
The projected energy can hence be evaluated using $\langle E_\textrm{Proj}\rangle=\langle E_\textrm{Numer}\rangle/ \langle N_\textrm{Denom}\rangle$.

Computing the stochastic matrix for an arbitrarily large system of determinants is computationally infeasible as the space scales as the power of the number of determinants, so we restrict ourselves to the simplest possible (interesting) system: two determinants, $a$ and $b$.  In this case the change on one determinant is independent of the change on the other, the elements of the stochastic matrix element are given by:
\begin{equation}
\Gamma_{\alpha, \beta} = p_{c  n_{a}, n^\prime_{a}} p_{c  n_{b}, n^\prime_{b}},
\end{equation}
where $p_{c  n_{a}, n^\prime_{a}}$ is the probability that the population on $a$ changes from $n_{a}$ to $n^\prime_{a}$ (\aref{trans_prob}).

We have constructed $\Gamma_{\alpha,\beta}$ for some simple systems and determined (by direct\cite{laug} or iterative\cite{Lehoucq97arpackusers} diagonalisation) the stationary distributions $\gamma_\alpha$ corresponding to the Perron-Frobenius eigenvector.  The Supplemental Material\cite{supp_mat} contains further examples of using the stationary distribution to examine the behaviour of the ensemble of FCIQMC states.

\section{Population Control Bias}
\label{finite_pop_bias}

\begin{figure*}
\includegraphics{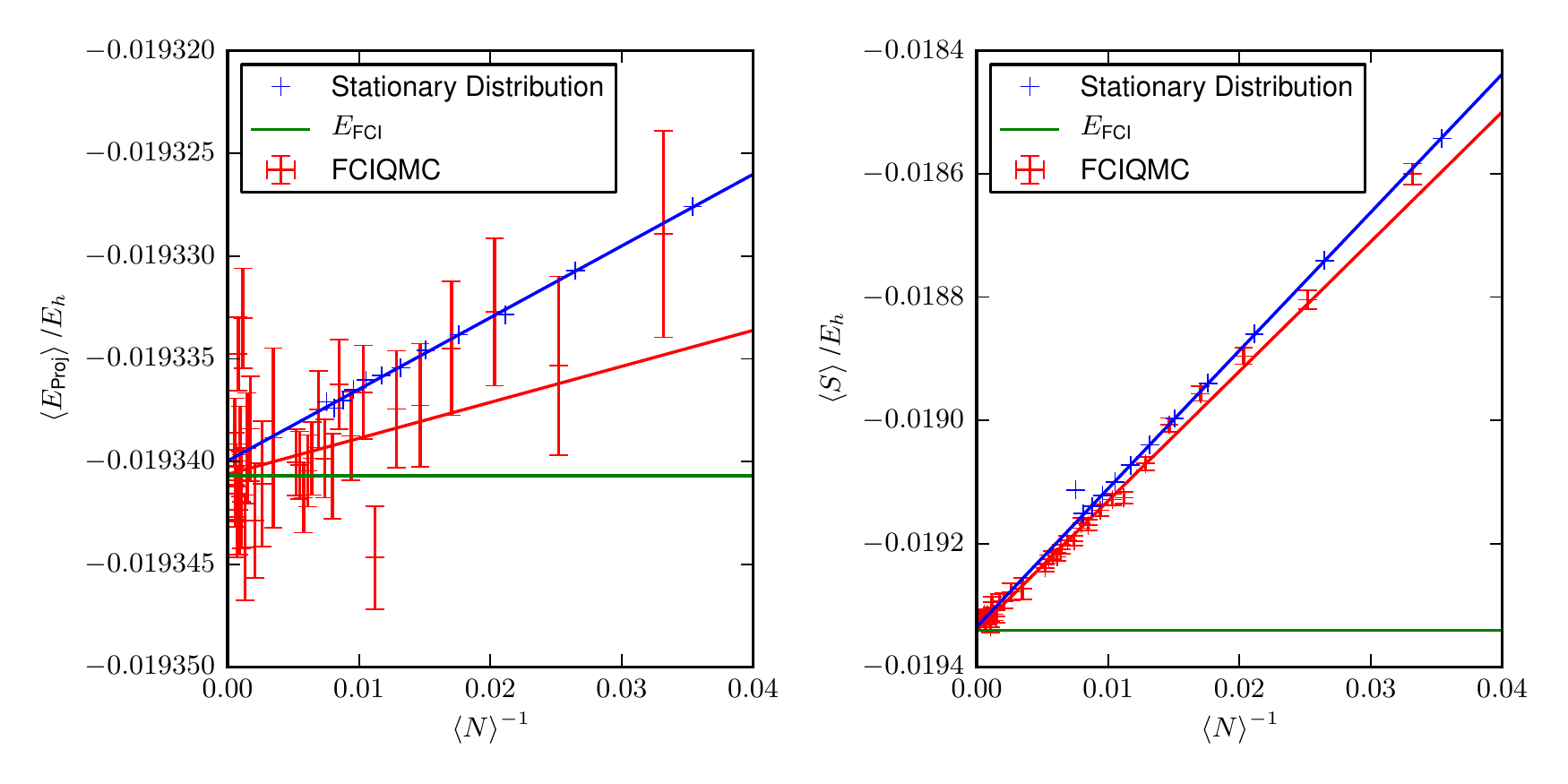}
\caption{\label{finite-pop-1} The energy estimators as a function of $1/\langle N\rangle$ for H$_2$ (STO-3G basis, internuclear separation $0.7122$\AA{}).  The exact means were calculated from the means of the stationary distributions.  Only states with up to 150 psips on each determinant were included in the transition matrix calculations.  The FCIQMC estimates were calculated from a single chain and the error estimated by blocking\cite{10.1063/1.457480}. The bias in both estimates of the correlation energy decays with the inverse of the average number of psips. Linear fits were performed with numpy\cite{:/content/aip/journal/cise/9/3/10.1109/MCSE.2007.58}. Errors were weighted in the fits for the FCIQMC data using the sum of the variance of $1/\langle N\rangle$ and variance of the energy estimator. The state with the smallest $1/\langle N\rangle$ was removed from the fit as the energy is too large to fit a linear slope. This is caused by the stochastic matrix being truncated at the state with $150$ psips on both determinants, states after this truncation have become important at this point.} 
\includegraphics{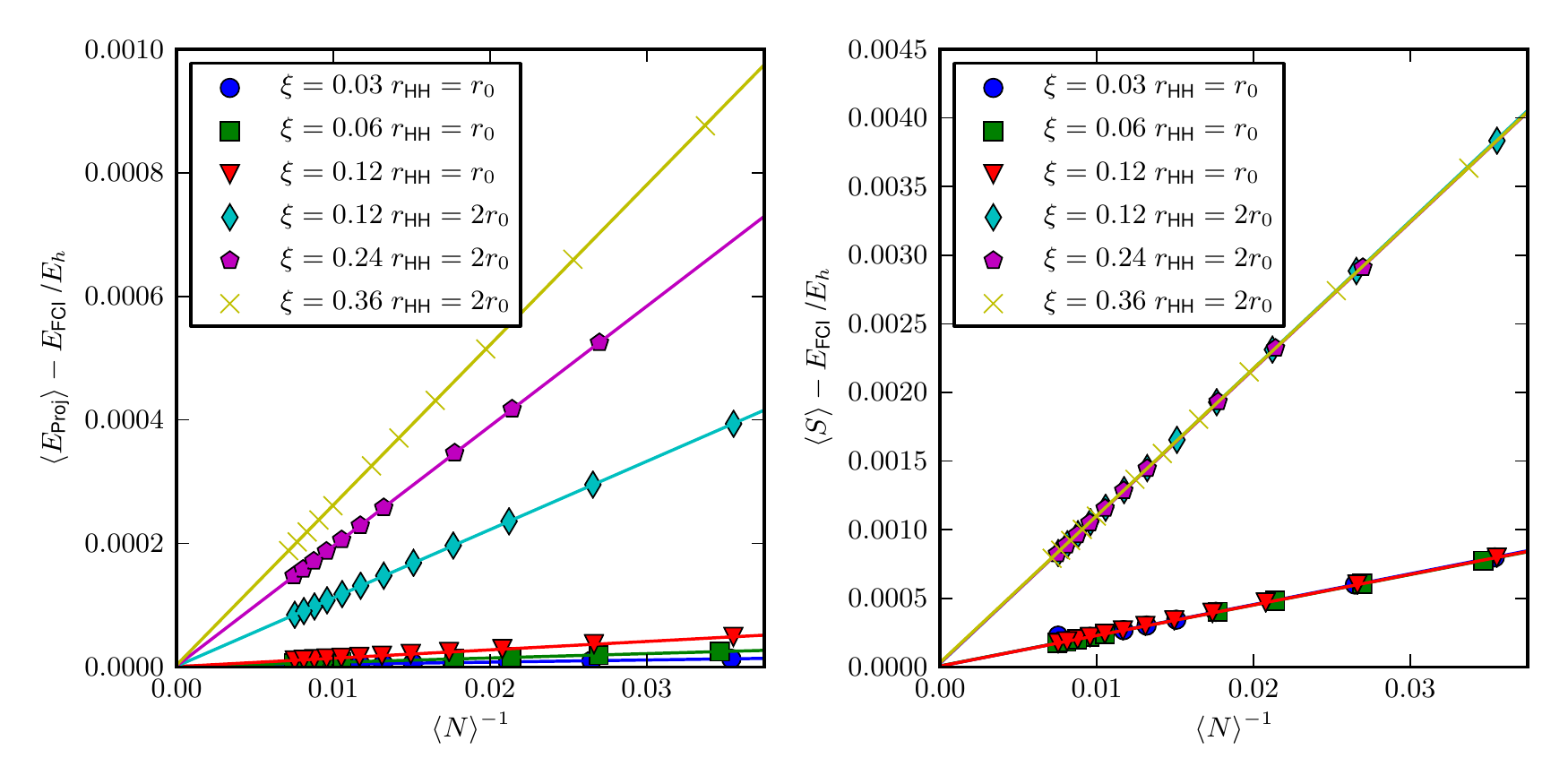}
\caption{\label{finite-pop-2} The energy estimators calculation from the means of the stationary distribution as a function of $1/\langle N\rangle$ for H$_2$ (STO-3G basis, internuclear separation $r_\textrm{HH}=r_0$ and $2r_0$, where $r_0=0.7122$\AA{}) for different values of $\xi$, the shift damping parameter.  Only states with up to 150 psips on each determinant were included in the transition matrix calculations.  The bias in the projected energy can be reduced by decreasing $\xi$ whereas the bias in the shift remains the same. Fits with $1/{\langle N \rangle}$ were performed with numpy\cite{:/content/aip/journal/cise/9/3/10.1109/MCSE.2007.58}. Again we remove the state with the smallest $1/\langle N\rangle$ for $\xi=0.03$ for $r_\textrm{HH}=r_0$.}
\end{figure*}

In order to achieve a finite population in a simulation, we must resort to population control by introducing a shift which itself is dependent upon the current total population.
However, this process introduces a feedback into the propagator and hence a systematic bias\cite{umrigar:2865}.  Random fluctuations causing the population to increase (i.e.\ the psip distribution enters a low energy region of phase space) or decrease (higher energy region of phase space) are moderated by a corresponding decrease or increase in the shift.  Both actions lead to an increase in the time-averaged energy estimators.

In DMC this bias is known\cite{umrigar:2865,PhysRevE.51.3679} to scale as $\langle N \rangle^{-1}$, though we know of no previous investigations of population control bias in FCIQMC\@.  We suspect that the effect has most likely been obscured by the stochastic error in all previous FCIQMC studies.
With the aid of exact energy estimators from the transition matrix, we are now able to investigate the magnitude of any bias present.
We feel it is important to understand where population control bias is likely to cause a problem if a small stochastic error is desired.
In addition to the energy estimators from the transition matrix, we shall also investigate them from single chains via blocking analyses of single FCIQMC calculations to compare both methods and use them to quantify the factors controlling population control bias.

\subsection{H$_2$ in a STO-3G Basis Set}

For different values of $N_s$, transition matrix and single chain calculations were performed on H$_2$ in a STO-3G basis at the equilibrium geometry of $0.7122 $ \AA{}\footnote{Hartree--Fock energy: $-1.1175058843$\,$E_h$.}, and the energy estimators evaluated.  \fref{finite-pop-1} shows the bias of the projected energy and shift estimators decreases with $1/\langle N\rangle$.
Though the single-chain calculations have relatively large stochastic errors, a similar bias in the energy and decay is also notable, and there is good agreement between the single chain and transition matrix results.
The fits for the transition matrix calculations, however, do not exactly intercept the $y$-axis at the correlation energy.   The worst extrapolation is for the projected energy, which disagrees by 2.3 $\pm$ 0.5\,$\mu E_h$. It is difficult to tell if this is caused by a loss of numerical precision, truncation in the transition matrix calculations or if there are higher order effects with small $\langle N\rangle$.

The prefactor in the $1/N$ scaling of the bias in the projected energy is affected by $\xi$, and damping less hard, i.e.\ decreasing $\xi$, reduces the prefactor (\fref{finite-pop-2}).
Population control bias also appears to be made worse in strongly correlated systems. \fref{finite-pop-2} shows population control bias as a function of $\xi$ for both H$_2$ in a STO-3G basis set at bond lengths of $0.7122 $\,\AA{} and $1.4244$\,\AA{}\footnote{Hartree--Fock energy: $-0.9338980552$\,$E_h$.}.
We may explain this by reviewing DMC, where, in the limit of a perfect trial function, there are no branching processes and thus no population control bias.
Equivalently in FCIQMC if there is no spawning there can be no population control bias.
 Although true only in the limit in which the Hilbert space is the set of eigenvectors of the Hamiltonian, this indicates that in the weakly correlated limit, there will be less population control bias, exactly as we observe.

\subsection{The Neon Atom in a cc-pVDZ basis}

\begin{figure*}[th]
\includegraphics{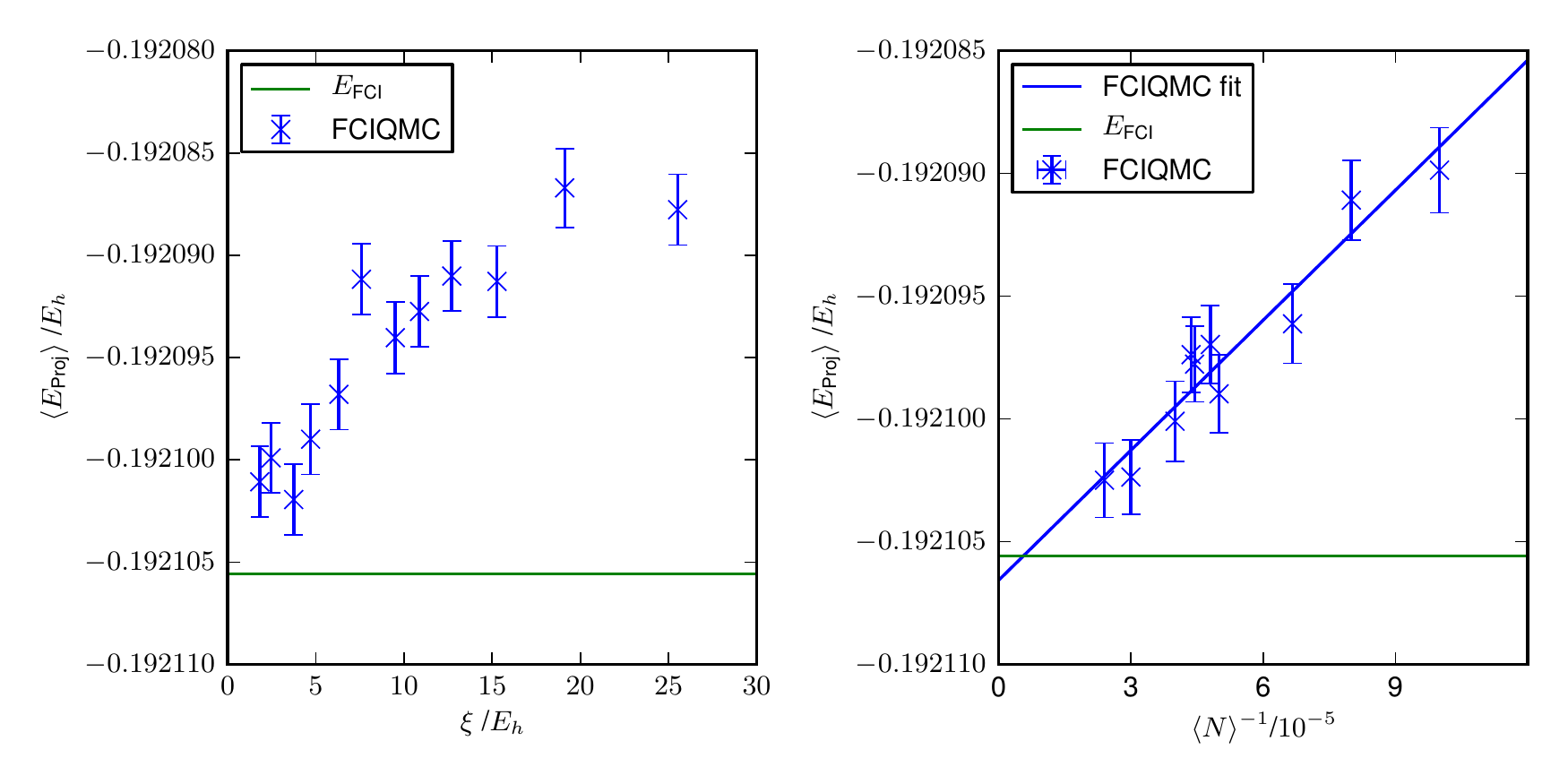}
\caption{\label{pop_bias_ne}
    Projected energy estimates from FCIQMC calculations on the Ne atom (cc-pVDZ basis set) as a function of (left) the shift damping parameter, $\xi$, (calculation details: $\delta \tau =0.005$, $20000000$ iterations where $\langle N\rangle$ has the value 10000) and (right) as a function of $1/\langle N \rangle$, where $N$ is the number of psips ($A=10$, $\xi=20$, $\delta \tau =0.005$, $200000000$ iterations or 72 hours on 12 cores whichever was shorter). Linear fits were performed with numpy.\cite{:/content/aip/journal/cise/9/3/10.1109/MCSE.2007.58}.
}
\end{figure*}

Population control bias is also potentially a significant source of a systematic error in systems which are large enough not to be trivially soluble (rendering transition matrix calculations computationally infeasible). We now turn to the neon atom in a cc-pVDZ basis\cite{10.10631.456153}\textsuperscript{,}\footnote{Hartree--Fock energy: $-128.4887755516$\,$E_h$.} which has a Hilbert space of $50000$ determinants.  This is small enough such that it is straightforward to compute the FCI energy via iterative diagonalisation but large enough such that most determinants have a small contribution to the wavefunction.  It was necessary to oversample the Hilbert space in H$_2$ (i.e.\ more psips than the number of determinants) whereas FCIQMC calculations in this neon system are stable with a significant undersampling of the space.  We shall investigate the effect of population control bias in this regime.

Changing the population control parameters affects both estimators of the correlation energy in the same way as H$_2$. \fref{pop_bias_ne}a shows the projected energy decaying towards the FCI energy as $\xi$ decreases until the estimator of the energy becomes within error bars. \fref{pop_bias_ne}b shows the bias in the projected energy decaying as $1/\langle N\rangle$. This intercepts the $y$-axis at -0.1921066(12)\,$E_h$ which is within errors of the FCI energy of -0.192105578\,$E_h$, suggesting we converge to the exact ground state as expected.   The population control bias is however significant and with about 10000 psips is about $20$\,$\mu E_h$.

As population control bias is clearly a problem, we sought to find a simple indicator of its magnitude in a calculation.  An investigation into the relationship between the variance of the shift and the magnitude of the population control bias did not yield any simple relationship.
With hindsight, a consideration of the causal relationship between the two makes it likely that not only the extent, but also the speed of variation of the shift is important. As such values are considerably more difficult to calculate, we will leave investigation of this connection to a future publication.

\begin{figure}
\includegraphics{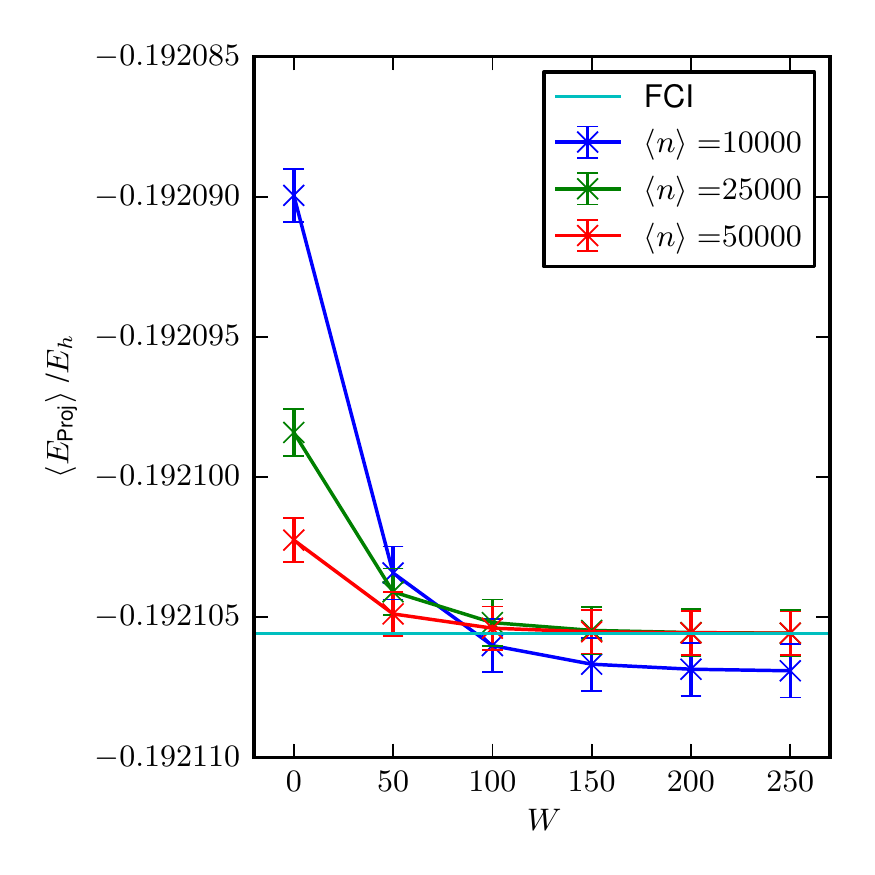}
\caption{\label{pop_bias_ne_renorm}
Reweighted projected energy estimate from FCIQMC calculations on the Ne atom (cc-pVDZ basis set) for different $\langle N \rangle$ as a function of $W$, the number of iterations reweighted over.  Four independent FCIQMC calculations were performed for each $\langle N\rangle$, and each point shows the mean of the reweighted $\langle E_\textrm{proj}\rangle$ for the four runs. }
\end{figure}

Instead, we have adopted a method used in DMC. The population bias can be both quantified and reduced by a reweighting technique based upon the history of the shift\cite{umrigar:2865}.  The contribution at a given time, $\tau$, to the numerator and denominator of the projected energy is weighted by taking into account the shift of the preceding $W$ iterations.  For $S_m$ denoting the shift $m$ iterations previously, the weight is given by:
\begin{equation}
w(\tau, W)  =\prod^W_{m=1} e^{-\delta \tau\left(S_m - \langle S \rangle \right)}.
\label{reweight}
\end{equation}
 The reweighting is implemented as a post-processing step on the output of a calculation.  The population control bias is effectively removed for sufficiently large $W$ at the cost of increasing the stochastic error and, as can be seen in \fref{pop_bias_ne_renorm}, the residual bias is of the order of the stochastic error bars. \footnote{FCIQMC projects out the ground state using $(1-(\hat{H} - S)) \delta \tau$ whereas $\exp(-\delta \tau (\hat{H} - S)) $ is used in DMC. This means weighting with \eref{reweight} is an approximation. The error introduced is second order in $\delta \tau$ (this can be shown using a Taylor expansion). The reweighting requires the projected energy and population at every individual time step.}

The value of $W$ ($\approx 250$) required for this procedure to converge is of the order of the serial correlation length.
We note that it is not possible to apply this method to the Markov Chain approach as the dependence of expectation values on calculation history makes the process non-Markovian.




\section{Discussion}
\label{discussion}

To summarise: we have demonstrated that FCIQMC is an example of Markov Chain Monte Carlo and computed the stochastic matrix for a two determinant system. Even though a two determinant system is the simplest non-trivial system, it still contains some of the inherent features of FCIQMC including population control bias.
A two determinant system can not have a sign problem unless the timestep is greater than the critical point. It would be interesting to extend these ideas to investigate the sign problem using a three determinant system, though the stochastic matrix may be inaccessible due to its scaling with the size of the Hilbert space.

Recently Petruzielo \emph{et al.}\ proposed to use floating point numbers to represent the population of psips on a determinant.\cite{PhysRevLett.109.230201} This adaption results in an uncountably infinite state space of the Markov chain. They also proposed to partition the determinant space into deterministic and stochastic subspaces, where the action of the Hamiltonian in the deterministic subspaces is applied exactly using sparse matrix multiplication and the action in the stochastic subspaces is sampled in the same way as in FCIQMC\@. Using floating point numbers as walker weights, as well as a multideterminental trial function significantly reduces the prefactor of the $1/\langle N \rangle$ scaling in population control bias.

The population control algorithm in DMC, as recommended in Ref.~\onlinecite{umrigar:2865}, is slightly different from that used in FCIQMC: the shift is updated from the `best current estimate' of the energy rather than from the previous value of the shift.  Using this population control algorithm would render FCIQMC non-Markovian.  Nonetheless we could use the stochastic matrix technique presented here to calculate the probability distribution of the shift in the limit of convergence of the projected energy. It would be interesting to investigate if this is a better method of population control for FCIQMC\@.

Using the population control approach given in Ref.~\onlinecite{booth:054106} (i.e.\ using \eref{shift-1} with $\gamma$ set in the region $0.01$ to $0.05$), may introduce population control bias, due to the factor of $\frac{1}{\delta \tau}$, if the timestep needed to converge a calculation needs to be small.
 This means population control bias is likely to be more of a problem for calculations which require smaller timesteps, such as strongly correlated systems, or calculations using coupled cluster Monte Carlo\cite{PhysRevLett.105.263004}.

We also note that converging FCIQMC calculations to $\mu E_h$ accuracy has previously been attempted.\cite{jcp/140/10/10.1063/1.4867383} In this regime, the population control bias could potentially become similar in magnitude to the stochastic error.

We recommend that one reweights the projected energy estimator, as suggested in Ref.~\onlinecite{umrigar:2865}, as it does not involve multiple expensive runs. If a large enough population is used the resultant estimate of the energy would be unbiased albeit with a larger stochastic error (though this has appeared negligibly larger in our tests).  Alternatively one should use a large population of psips and set $\xi$ to be as small as possible, such that the number of psips does not drop below the system-dependent critical population.  Doubling the number of psips in a simulation increases the equilibration time and possibly also the memory requirements. It is also important to perform enough steps to get an accurate estimate of the error.  
In choosing an appropriate value of $\xi$ there is a compromise to be made;  it is tempting to increase $\xi$ because it reduces the fluctuations in the total number of psips and, for larger systems, this can reduce the maximum amount of memory used during the calculation. However too large a $\xi$ will cause population control bias to become significant.

In conclusion, we caution users of FCIQMC and related methods to be aware that population control can introduce a significant bias in calculated energies.  We recommend that post-processing reweighting is used to quantify its magnitude  and the psip population and damping parameters be modified as suggested in this paper if needed.

\begin{acknowledgments}
The authors thank C.J.~Umrigar for several enlightening discussions, facilitated by the unique setting of The Towler Institute.
Calculations were performed using the Imperial College High Performance Computing Service\cite{ICHPC}  and figures were plotted using matplotlib.\cite{Hunter:2007}  JSS acknowledges the research environment provided by the Thomas Young Centre under Grant No.~TYC-101.  WAV is grateful to EPSRC for a studentship and AJWT thanks Imperial College for a Junior Research Fellowship and the Royal Society for a University Research Fellowship.
\end{acknowledgments}

\appendix
\section{Transition probabilities for a two determinant system}
\label{trans_prob}

Consider two determinants, $a$ and $b$, with states $\alpha$ and $\beta$ representing two states with a given signed number of psips on each determinant:
\begin{equation}
\begin{aligned}
    \alpha &= (n_{a}, n_{b}) \\
    \beta &= (n^\prime_{a}, n^\prime_{b}).
\end{aligned}
\end{equation}
Each psip independently attempts to spawn and die every timestep. The probability that $n$ psips succeed out of $N$ attempts is given by the probability mass function of the binomial distribution, $B(n, N, p) = \binom{N}{n} p^{n} {(1 - p)}^{N -  n}$, where $p$ is the probability of one psip spawning or dying independently and can be obtained from the FCIQMC algorithm.\cite{booth:054106}
The change on determinant $a$ is\cite{psip_prob_assumption}:
\begin{equation}
n^\prime_{a} -  n_{a} = -  \sgn(H_{ba}) \sgn(n_{b} ) n_{s a} - \sgn(H_{aa}-S) \sgn(n_{a})  n_{d  a}
\label{change}
\end{equation}
where $n_{s a}$ ($n_{d a}$) is the number of psips spawning onto (dying on) $a$.
As the spawning and death events on each determinant are independent, the probability $p_{c n_{a}, n^\prime_{a}}$ that the number of psips on $a$ changes from $n_{a}$ to $n^\prime_{a}$ via any possible combination of spawning and death is given by:
\begin{equation}
p_{c n_{a}, n^\prime_{a}} = \sum_{n_{s a}} B(n_{s a}, n_{b}, P_s(a|b)) B(n_{d  a}, n_{a}, P_d ( a)),
\end{equation}
where $P_s(a|b)$ is the probability that a psip on $a$ spawns a child onto $b$ and $n_{d  a}$ is the probability that a psip died on $a$.

\bibliography{paper}

\begin{thebibliography}{35}
\bibitem [{psi()}]{psip_prob_assumption}%
  \BibitemOpen
  \href@noop {} {}\bibinfo {note} {{We assume that the independent psip
  probabilities are less than or equal to one for simplicity; the
  generalisation is straightforward.}}\BibitemShut {Stop}%
\end{thebibliography}

\end{document}